\documentclass[10pt,a4paper]{article}
\usepackage{amsmath,amssymb}
\usepackage{graphicx}
\newcommand{\sech}{\operatorname{sech}}
\newcommand{\I}{\ensuremath{\mathrm{i}\,}}
\newcommand{\E}{\ensuremath{\mathrm{e}\,}}
\title{%
{\vspace{-3cm} \normalsize \hfill\parbox{38mm}{MS-TP-08-01}}\\[20mm]
Interfacial roughening in field theory}

\author{Michael H.~K\"opf and Gernot M\"unster%
\thanks{Institut f\"ur Theoretische Physik,
        Universit\"at M\"unster,
        Wilhelm-Klemm-Str.~9, \mbox{D-48149 M\"unster,} Germany;
        e-mail: munsteg@uni-muenster.de}}
\date{January 23, 2008}
\begin{document}
\maketitle

\begin{abstract}
In the rough phase, the width of interfaces separating different phases of
statistical systems increases logarithmically with the system size. This
phenomenon is commonly described in terms of the capillary wave model, which
deals with fluctuating, infinitely thin membranes, requiring ad hoc cut-offs
in momentum space. We investigate the interface roughening from first
principles in the framework of the Landau-Ginzburg model, that is
renormalized field theory, in the one-loop approximation. The interface
profile and width are calculated analytically, resulting in finite
expressions with definite coefficients. They are valid in the scaling region
and depend on the known renormalized coupling constant.\\[5mm]
\textbf{KEY WORDS}: Interfaces, field theory
\end{abstract}
%=======================================================================
\section{Introduction}

Interface roughening is a phenomenon which has attracted interest of
experimental and theoretical investigators, see
e.g.~\cite{RW82,HW69,La92,Th04,WSMB97,WSMB99,MC}, since its discovery
\cite{BLS65}. It is displayed by interfaces, separating different coexisting
phases or substances of a system of statistical physics, in a range of
temperatures $T_R < T < T_c$ between the roughening temperature $T_R$ and
the critical temperature $T_c$. Roughening manifests itself in a
characteristic dependence of the interface width on the system size. For an
interface of diameter $L$ the width increases logarithmically with $L$ in
the rough phase, whereas it remains constant of the order of the correlation
length $\xi$ for temperatures below $T_R$.

This effect is commonly described theoretically in terms of the capillary
wave model or drumhead model \cite{BLS65}. In this model the interface is
represented in an idealized way by an infinitely thin fluctuating membrane,
so that the instantaneous microscopic interface profile is a sharp step
function between the two phases. Nevertheless, in the thermal average the
capillary wave fluctuations produce a continuous density profile with a
finite width $w$, which can be shown to be given by an integral over all
wave-numbers of the fluctuations, which is essentially of the form
\begin{equation}
\label{cwintegral}
w^2 = \frac{1}{2\pi\sigma}
\int_{k_{\textrm{\scriptsize min}}}^{k_{\textrm{\scriptsize max}}} dk\ k^{D-4}\,,
\end{equation}
where $D$ is the number of dimensions of space and $\sigma$ is the interface
tension. A natural lower limit on the wave numbers is given by the system
size,
\begin{equation}
k_{\textrm{\scriptsize min}} = \frac{\textrm{const.}}{L}\,.
\end{equation}
In order to avoid the divergence of the integral, an upper cut-off
$k_{\textrm{\scriptsize max}}$ has to be introduced. As there should be no
waves with wavelength smaller than the intrinsic width of the physical
interface, the upper cut-off is taken to be of the order of the inverse
correlation length. In the case of $D=3$, considered here, one obtains
\begin{equation}
\label{cwwidth}
w^2 = \frac{1}{2\pi\sigma} \ln \frac{L}{c\, \xi}
\end{equation}
with an unknown constant $c$. The logarithmic increase with $L$ is due to
the contribution of capillary waves with long wavelengths near the system
size $L$.

Complementary to the capillary wave model is the mean field description of
interfaces. In mean field theory and its field theoretic refinements,
interfaces possess an intrinsic continuous profile with a well-defined
width, which is proportional to the bulk correlation length and does not
depend on the system size.

Mean field and capillary wave theory can be combined in the ``convolution
approximation'' \cite{Ja84,Wi72}.  In this picture the intrinsic profile
describes the interface on a microscopic scale of the order of the
correlation length, while capillary wave theory describes the macroscopic
interface fluctuations of wavelengths much larger than the correlation
length. The intrinsic profile is thus centered around a two-dimensional
surface subject to capillary wave fluctuations. In the convolution
approximation the square of the resulting total interface width is obtained
as a sum of the intrinsic part and the capillary wave contribution,
\begin{equation}
\label{cawidth}
w^2 = c_1 \xi^2 + \frac{1}{2\pi\sigma} \ln \frac{L}{c_2\, \xi}\,.
\end{equation}

The description of rough interfaces by means of the capillary wave model and
the convolution approximation is unsatisfactory for different reasons. First
of all, it has so far not been possible to define the concept of an
intrinsic interface profile and width unambiguously outside a given theory. 
In experiments or Monte Carlo simulations of systems with interfaces, the
observed interface profile and width are the total ones, including the
effects of the intrinsic structure as well as of the capillary waves, and
there is no clear way to separate the intrinsic structure from the effects
of capillary waves. Secondly, the models sketched above contain ad hoc
constants, whose numerical values are arbitrary and cannot be fixed
unambiguously within the models.

In this article we investigate the profile and width of rough interfaces in
a coherent approach from first principles. Statistical systems with
coexisting phases, separated by interfaces, are described in the framework
of the field theoretic version of the Landau-Ginzburg model, including
fluctuations on all length scales. No cut-off on wave-numbers is introduced.
For explicit calculations we employ the one-loop approximation. It should be
noted, however, that an extension to arbitrary higher loop orders is
possible in principle. The interface profile, resulting from the
calculation, shows the expected logarithmic broadening with the system size
$L$. We obtain analytical results for the numerical coefficients, which are
fixed unambiguously in this approach.

Interfaces have been studied before in the framework of field theory by
other authors. In \cite{OK77,RJ78} the profile is calculated to first order
in the $\epsilon$-expansion, where $D=4-\epsilon$ and an extrapolation to
$\epsilon=1$ is necessary. The $\epsilon$-expansion is an expansion around
the four-dimensional case. As can be seen from Eq.~(\ref{cwintegral}), in
four dimensions the contribution of long-wavelength modes converges and no
roughening is present. This has the consequence that within the
$\epsilon$-expansion, even after extrapolation to $D=3$ dimensions,
roughening effects do not show up, as is well known. The calculation of
\cite{OK77} is extended to include the effects of an external field in
\cite{SL86}.

Our calculations are performed in $D=3$ physical dimensions in contrast to
the $\epsilon$-expansion. The three-dimensional approach is based on a
systematic expansion in a dimensionless coupling \cite{Parisi80,LZ80}.
Ultraviolet divergences are treated by dimensional regularization
($D=3-\epsilon$), which does not vitiate the fact that the results for
physical quantities strictly refer to $D=3$ dimensions. This is also seen
explicitly by the fact that the calculation reveals the typical roughening
effects. Renormalization of the three-dimensional field theory is performed
in the scheme used in \cite{Mu90} to two-loop order, employing the results
of \cite{MH94,GKM96}.

A three-dimensional study has previously been done in \cite{JR78}, where the
interface profile is considered in $D=3$ dimensions at one-loop order in the
presence of an external gravitational field. A functional form of the
profile is given, including capillary wave effects. The dependence on the
system size is, however, not considered. We shall compare our results with
the ones of \cite{JR78} below.

%=======================================================================
\section{Interfaces in field theory}

In the framework of field theory, the system under consideration, possessing
interfaces, is described by an order parameter field $\phi(x)$ representing
the difference between the concentrations of the two coexisting phases. The
physics of the system is governed by the Landau-Ginzburg 
Hamiltonian~\cite{Bellac}
\begin{equation}
H[\phi] = \int\!\!d^3x\,\mathcal{H}(\phi(x))
\end{equation}
with the Hamiltonian density
\begin{equation}
\label{Ham}
\mathcal{H}(\phi) = \frac{1}{2} \partial_\mu  \phi \partial_\mu  \phi +
V_0(\phi)\,.
\end{equation}
In the situation with interfaces the potential is of the double-well type,
\begin{equation}
V_0(\phi) = \frac{g_0}{4!}\left(\phi^2 -v_0^2\right)^2.
\end{equation}

Mean field theory amounts to the classical approximation where fluctuations
are neglected. The minima of the potential then correspond to the two
homogeneous phases. The mean field correlation length $\xi_0$ is defined
through the second moment of the correlation function in the mean field
approximation. It is given by the second derivative of the potential in its
minima:
\begin{equation}
\xi_0^2 = \left( V''_0(v_0) \right)^{-1} = \frac{3}{g_0 v_0^2}\,.
\end{equation}
With the bare mass $m_0$, defined by
\begin{equation}
m_0 = \frac{1}{\xi_0},
\end{equation}
the Hamiltonian density can be written as
\begin{equation}
\mathcal{H}(\phi) = \frac{1}{2} \partial_\mu  \phi \partial_\mu  \phi
- \frac{m_0^2}{4} \phi^2 + \frac{g_0}{4!} \phi^4
+ \frac{3}{8} \frac{m_0^4}{g_0}\,.
\end{equation}

The simplest description of interfaces is also based on mean field theory
\cite{vdW93}. In this approximation the interface profile is given by
minimization of the Hamiltonian $H$ with boundary conditions appropriate for
an interface.  The corresponding field equation
\begin{equation}
\frac{\delta H}{\delta \phi(x)} = 0
\end{equation}
leads to the differential equation
\begin{equation}
\Delta \phi - V'_0(\phi) = 0\,.
\end{equation}
If we choose the interface to be perpendicular to the $z$-axis, we find
the typical hyperbolic tangent profile \cite{CH58}
\begin{equation}
\phi_0^{(z_0)}(z) = v_0 \tanh \left(\frac{z-z_0}{2 \xi_0}\right).
\end{equation}
Its width is proportional to the mean field correlation length $\xi_0$. The
parameter $z_0$ specifies the location of the interface.

Essential for a field theoretic treatment, as being considered in this
article, are corrections to mean field theory coming from fluctuations of
the order parameter field. They can be calculated systematically in
renormalized perturbation theory. The fluctuations result in different
modifications of the mean field result, as will be considered in detail
below. First of all, higher order corrections change the form of the profile
from the tanh-function to a different function. Secondly, renormalization of
the parameters $v_0$ and $\xi_0$ becomes necessary and, as a result, the
mean field correlation length $\xi_0$ is replaced by the physical
correlation length $\xi$, which diverges near the critical point with a
characteristic exponent $\nu$. Finally, long-wavelength fluctuations lead to
the roughening phenomenon, which implies a broadening of the interface, such
that its width depends logarithmically on the system size and diverges in
the limit of an infinite system.

The partition function for the system with an interface can be written as a
functional integral of the form
\begin{equation}
Z = \int\!\!\mathcal{D}\varphi\,\exp (- H[\phi_0 + \varphi])\,,
\end{equation}
where $\phi_0(z)$ is a classical interface solution as given above and
$\varphi(x)$ denotes the fluctuations around it. The Hamiltonian density,
expressed in terms of $\varphi$, reads
\begin{equation}
\label{Hamfluct}
\mathcal{H}(\phi_0 + \varphi) = 
\mathcal{H}(\phi_0) + \frac{1}{2} \varphi(x) K \varphi(x)
+ \frac{g_0}{3!} \phi_0(x) \varphi^3(x)
+ \frac{g_0}{4!} \varphi^4(x)\,.
\end{equation}
Here the operator $K$ is given by
\begin{equation}
K = - \Delta - \frac{m_0^2}{2} + \frac{g_0}{2} \phi_0^2(x)\,,
\end{equation}
where $\Delta$ is the Laplacean.

In the loop expansion the quadratic terms in $\mathcal{H}$ are treated by
means of Gaussian functional integrals, and the higher order terms are taken
into account by Taylor expansions.

The spectrum of $K$ is known analytically \cite{R75}. We have to employ it
for our calculation and give details below. At this point we would like to
draw the attention to the fact that $K$ has a single zero mode
\begin{equation}
K \psi(x) = 0\,.
\end{equation}
The zero mode of the fluctuation operator is directly related to
translations of the interface, as parameterized by the parameter $z_0$. For
every value of this parameter, the function $\phi_0^{(z_0)}$ is a solution
of the classical field equation. This implies that
\begin{equation}
\psi(z) = \frac{d \phi_0^{(z_0)}(z)}{d z_0}
\end{equation}
is a zero mode of $K$.

The existence of a zero mode requires to treat the corresponding
fluctuations, which are proportional to $\psi$, separately from the
remaining Gaussian integrals in the functional integrals. This is done by
the method of collective coordinates \cite{GS75}. The collective coordinate
in question is $z_0$. In the Gaussian integral it is set to an arbitrary
value, which we choose to be $z_0=0$, and the fluctuations are restricted to
the space $\mathcal{N}_{\perp}$ of functions orthogonal to the zero mode
$\psi$:
\begin{equation}
\int\!\!d^3x\, \varphi(x) \psi(x) = 0\,.
\end{equation}
When expectation values in the presence of an interface are calculated,
integration over $z_0$ would imply averaging over all translations of the
interface, leading to translationally invariant results. In case of the
interface profile, however, this is obviously not appropriate, since one is
interested in the profile function relative to the position of the
interface. Therefore integration over $z_0$ has to be omitted, leaving us
with Gaussian integrals over $\mathcal{N}_{\perp}$. So the interface profile
is given by
\begin{equation}
\phi_c (x) = \phi_0 (x) + \phi_f(x)
\end{equation}
with
\begin{equation}
\phi_f(x) = \langle \varphi(x) \rangle =
\frac{1}{Z'}
\int_{\mathcal{N}_{\perp}}\!\!\mathcal{D}\varphi\,\varphi(x)\,
\exp (- H[\phi_0 + \varphi])\,.
\end{equation}

%=======================================================================
\section{The profile equation}

For functional integrals over the fluctuation field $\varphi \in
\mathcal{N}_{\perp}$ Feynman rules can be set up analogously
to the usual case. The propagator and vertices can be read off the
Hamiltonian (\ref{Hamfluct}). The propagator is the inverse of the
fluctuation operator restricted to $\mathcal{N}_{\perp}$:
\begin{equation}
K' = K |_{\mathcal{N}_{\perp}}.
\end{equation}
There are three-point and four-point vertices, given by
$$
\raisebox{-6.2mm}{\includegraphics[scale=0.4]{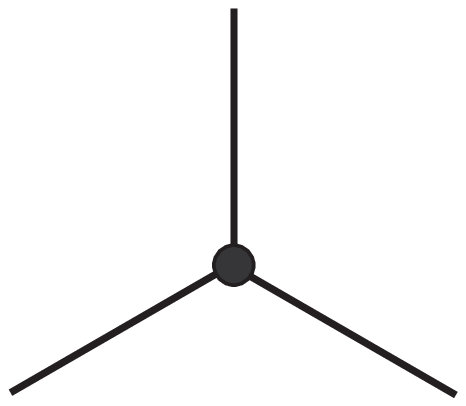}} 
= -g_0\phi_0(z)\,, \hspace{5mm}
\raisebox{-9mm}{\includegraphics[scale=0.4]{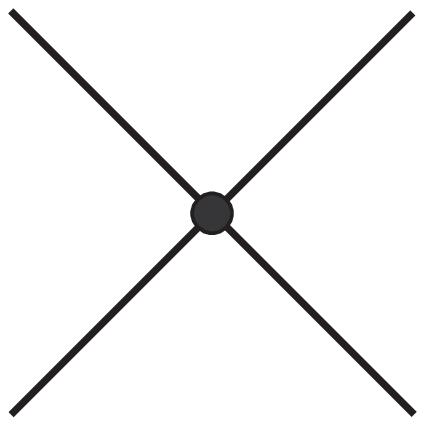}}
= -g_0\,.
$$

The fluctuation part the interface profile gets contributions from all
orders of the loop expansion:
\begin{equation}
\phi_f(x) = \phi_1(x) + \phi_2(x) + \dots
\end{equation}
In the one-loop approximation, which we employ, the Feynman diagram
contributing to the profile function
$$
\raisebox{-9mm}{\includegraphics[scale=0.4]{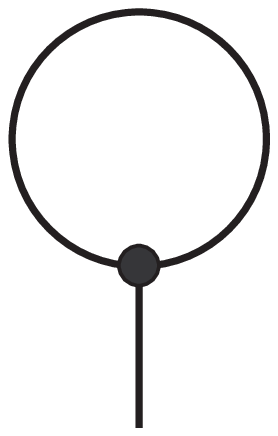}}
$$
leads to
\begin{equation}
\label{phi1}
\phi_1(x) = - \frac{g_0}{2} 
\int\!\!d^3x'\,K'^{-1}(x,x')\,K'^{-1}(x',x') \phi_0(x')\,.
\end{equation}
Here the kernel of the inverse operator $K'^{-1}$ enters. It would be
possible to calculate $\phi_1$ from this expression. It is, however, more
convenient to obtain it as a solution of a differential equation. Acting
with the operator $K$ on Eq.~(\ref{phi1}), we obtain the profile equation
\begin{equation}
\label{profeq}
K \phi_1(x) + \frac{g_0}{2} K'^{-1}(x,x) \phi_0(x) = 0\,.
\end{equation}
In order to solve this equation we need the explicit form of $K'^{-1}(x,x)$,
which is discussed below.

An alternative derivation of the profile equation is based on the so-called
effective action $\Gamma[\Phi]$, which is a functional of a field $\Phi(x)$.
$\Gamma[\Phi]$ is obtained by Legendre transformation from the free energy in the
presence of a non-constant external field. For constant $\Phi$ the effective
action reduces to the Gibbs potential. For a definition and discussion see
e.g.\ \cite{KM07}. Calculating $\Gamma[\Phi]$ in the one-loop approximation
and finding the interface profile $\phi(x)$ as a stationary point of
$\Gamma$,
\begin{equation}
\frac{\delta \Gamma}{\delta \Phi(x)} = 0\,,
\end{equation}
again leads to Eq.~(\ref{profeq}).

%=======================================================================
\section{Solution of the profile equation}

The inverse of the fluctuation operator $K'$ at coinciding arguments, which
enters the profile equation, can be obtained by means of the spectral
representation. $K$ is the sum of the negative two-dimensional Laplacean and
a one-dimensional Schr\"odinger operator $\tilde{K}$,
\begin{equation}
K = - \Delta^{(2)} + \tilde{K}\,,
\end{equation}
where
\begin{equation}
\tilde{K} = - \partial_z^2 + m_0^2 
- \frac{3 m_0^2}{2} \sech^2 \left(\frac{m_0}{2} z \right).
\end{equation}
The negative Laplacean on the $L \times L$ square has eigenvalues
\begin{equation}
k^2 \quad \textrm{with} \quad 
\vec{k}=\frac{2\pi}{L}\vec{n}, \quad \vec{n} \in \mathbb{Z}^{2},
\end{equation}
and corresponding eigenfunctions
\begin{equation}
\varphi_{\vec{n}}(\vec{x}) 
= L^{-1}\ \E^{\I\frac{2\pi}{L}\vec{n}\cdot\vec{x}}, 
\quad \vec{x}\in[0,L]^2\,.
\end{equation}
The spectrum of $\tilde{K}$ is known exactly \cite{R75}. It consists of two
discrete eigenvalues
\begin{equation}
\omega^{(0)} = 0,\:\: 
\psi_{0}(z) = \sqrt{\frac{3m_0}{8}} \sech^2\left(\frac{m_0}{2}z\right),
\end{equation}
\begin{equation}
\omega^{(1)} = \frac{3}{4}m_0^2,\:\: 
\psi_{1}(z) = \sqrt{\frac{3m_0}{4}} \tanh\left(\frac{m_0}{2}z\right)
\sech\left(\frac{m_0}{2}z\right),
\end{equation}
and a continuum
\begin{equation}
\omega_p = m_0^2 + p^2 \quad \mathrm{with} \:\; p \in \mathbb{R}, 
\end{equation}
\begin{equation}
\psi_{\omega_p}(z) = \mathcal{N}_p \E^{\I pz} \left[ 2p^2 +
  \frac{m_0^2}{2} - \frac{3}{2}m_0^2\tanh^2\left( \frac{m_0}{2}z \right)
  + 3\I m_0 p \tanh\left( \frac{m_0}{2}z \right) \right]
\end{equation}
with the normalization factor
\begin{equation}
\mathcal{N}_p = (2\pi (4p^4 + 5m_0^2\,p^2 + m_0^4))^{-\frac{1}{2}}.
\end{equation}
The spectrum of $K$ is thus given by
\begin{equation}
\lambda_{\vec{n}\omega} = \frac{4\pi^2}{L^2} n^2 + \omega\,,
\qquad
\Psi_{\vec{n}\omega}(x) = \varphi_{\vec{n}}(\vec{x})\, \psi_\omega(z),
\end{equation}
where $\omega$ runs through the eigenvalues of $\tilde{K}$. The zero mode,
discussed above, is represented by $\Psi_{\vec{0} 0}$.

In terms of the spectrum we write
\begin{equation}
K'^{-1}(x,x) = 
\underset{\lambda}{{\displaystyle\sum\hspace{-13pt}\int \,}}
\psi_\lambda(x) \psi^*_\lambda(x) \frac{1}{\lambda}.
\end{equation}
Inserting the explicit expressions, we obtain
\begin{equation}
K'^{-1}(x,x) = 
C_0 + (C_1 - C_2 + C_4)\sech^4\left(\frac{m_0}{2}z\right) + 
(C_2 + C_3)\sech^2\left(\frac{m_0}{2}z\right),
\end{equation}
where the coefficients $C_i$ are
\begin{align}
C_0 & = \frac{1}{2 \pi}\int\!\!dp\,\sum_{\vec{n}}
        \frac{1}{4\pi^2n^2+(m_0^2+p^2)L^2}\,, \\
C_1 & = \frac{3 m_0}{8} \sum_{\vec{n} \ne \vec{0}} \frac{1}{4\pi^2n^2}\,, \\
C_2 & = \frac{3 m_0}{4} \sum_{\vec{n}} 
        \frac{1}{4\pi^2 n^2 + \frac{3}{4}m_0^2L^2}\,, \\
C_3 & = - 3 m_0^2 \int\!\!dp\,\mathcal{N}^2_p \sum_{\vec{n}} 
        \frac{m_0^2+p^2}{4\pi^2n^2 + (m_0^2 + p^2)L^2}\,, \\
C_4 & = \frac{9}{4} m_0^4 \int\!\!dp\,\mathcal{N}^2_p \sum_{\vec{n}} 
        \frac{1}{4\pi^2n^2 + (m_0^2 + p^2)L^2}.
\end{align}
These expressions are divergent and have to be regularized, as discussed
below.

With the explicit form of $K'^{-1}(x,x)$ at hand, the solution of the
profile equation is found as
\begin{align}
\phi_1(z) = \
&\frac{g_0 v_0}{2 m_0^2} \bigg\{
C_0 \tanh\left( \frac{m_0}{2}z \right) \nonumber\\
&-\bigg[ \frac{2}{3}(C_1-C_2+C_4) \tanh\left(\frac{m_0}{2}z\right) \nonumber\\
&- (C_0+C_2+C_3) \frac{m_0}{2}z \bigg] \sech^2\left(\frac{m_0}{2}z\right)
\bigg\}.
\end{align}
Written in this way, the expression for the profile contains the divergent
coefficients $C_i$ as well as the bare parameters $g_0$, $m_0$ and $v_0$. In
order to arrive at a finite expression in terms of physical parameters,
renormalization has to be performed.

The divergences have to be treated in some regularization scheme. We choose
to employ dimensional regularization in $D=3-\epsilon$ dimensions. It should
be noted that this does not amount to an $\epsilon$-expansion, since after
renormalization $\epsilon$ is sent to zero, whereas in the
$\epsilon$-expansion one has $D=4-\epsilon$ and the results have to be
extrapolated to $\epsilon = 1$. So our use of dimensional regularization
does not vitiate the fact that the results for physical quantities strictly
refer to $D=3$ dimensions. Using other regularization schemes, like
Pauli-Villars, would lead to the same final results.

We adopt the renormalization scheme used in \cite{Mu90} to one-loop order.
The renormalized mass $m_R = 1 / \xi$ is equal to the inverse correlation
length $\xi$, which in turn is defined through the second moment of the
correlation function. The field $\phi$ and its expectation value $v$ are
renormalized according to
\begin{equation}
\phi_R(x) = \frac{1}{\sqrt{Z_R}}\,\phi(x)\,, \qquad
v_R = \frac{1}{\sqrt{Z_R}}\,v\,,
\end{equation}
where $Z_R$ is the usual field renormalization factor. The renormalized
coupling is specified as in \cite{LW87} through
\begin{equation}
g_R = \frac{3 m_R^2}{v_R^2}\,.
\end{equation}
In addition we define a dimensionless renormalized coupling according to
\begin{equation}
u_R = \frac{g_R}{m_R^{4-D}}\,.
\end{equation}
Employing the relations given in \cite{MH94,GKM96}, the bare quantities
$m_0$ and $g_0$ are expressed in terms of their renormalized counterparts.

The coefficients $C_i$ are evaluated in the same scheme. Leaving out the
lengthy details, we quote the results
\begin{align}
C_0 &= - \frac{m_0}{4\pi} \\
C_2 + C_3 &= \frac{3 m_0}{16\pi} \ln 3 \\
C_1 - C_2 + C_4 &= \frac{3 m_0}{16\pi} (-\alpha + \ln(m_0 L))
\end{align}
with
\begin{equation}
\alpha =
\ln\left( \frac{3\Gamma^2(1/4)}{2\sqrt{\pi}} \right) - \gamma
\approx 1.832\,,
\end{equation}
where $\gamma \approx 0.577$ is Euler's constant.

The coefficients $C_i$ contain additional terms decaying exponentially fast
with $L$, which we neglect here.

Inserting everything into the expression for the interface profile yields
the renormalized interface profile $\phi_R(z)$ depending on the parameters
$m_R$ and $u_R$. In this expression the divergences are cancelled, as they
should. Expanding consistently in powers of $u_R$ up to the first order, we
obtain
\begin{align}
\label{profile}
\phi_R(z) = v_R \bigg\{
&\tanh(\frac{m_R}{2}z) \nonumber\\
&+ \frac{u_R}{16\pi} (\alpha - \ln(m_R L) ) 
\tanh(\frac{m_R}{2}z) \sech^2(\frac{m_R}{2}z) \nonumber\\
&- \frac{u_R}{32\pi} \left( 3 \ln 3 - \frac{13}{4} \right)
\frac{m_R}{2}z \sech^2(\frac{m_R}{2}z)
\bigg\},
\end{align}
which is the central result of this article. Asymptotically, for $|z|
\rightarrow \infty$, the profile approaches the bulk expectation value
$\pm v_R$, corresponding to the pure phases of the system, as it should be.
Note that the profile depends logarithmically on the system size $L$,
revealing the effect of capillary wave fluctuations. It is this term,
depending on $m_R L$, which represents the deviation from the Fisk-Widom
\cite{FW60} scaling form $\phi(m_R z)$.

\begin{figure}[hbt]
\vspace{.8cm}
\centering
\includegraphics[width=9.5cm]{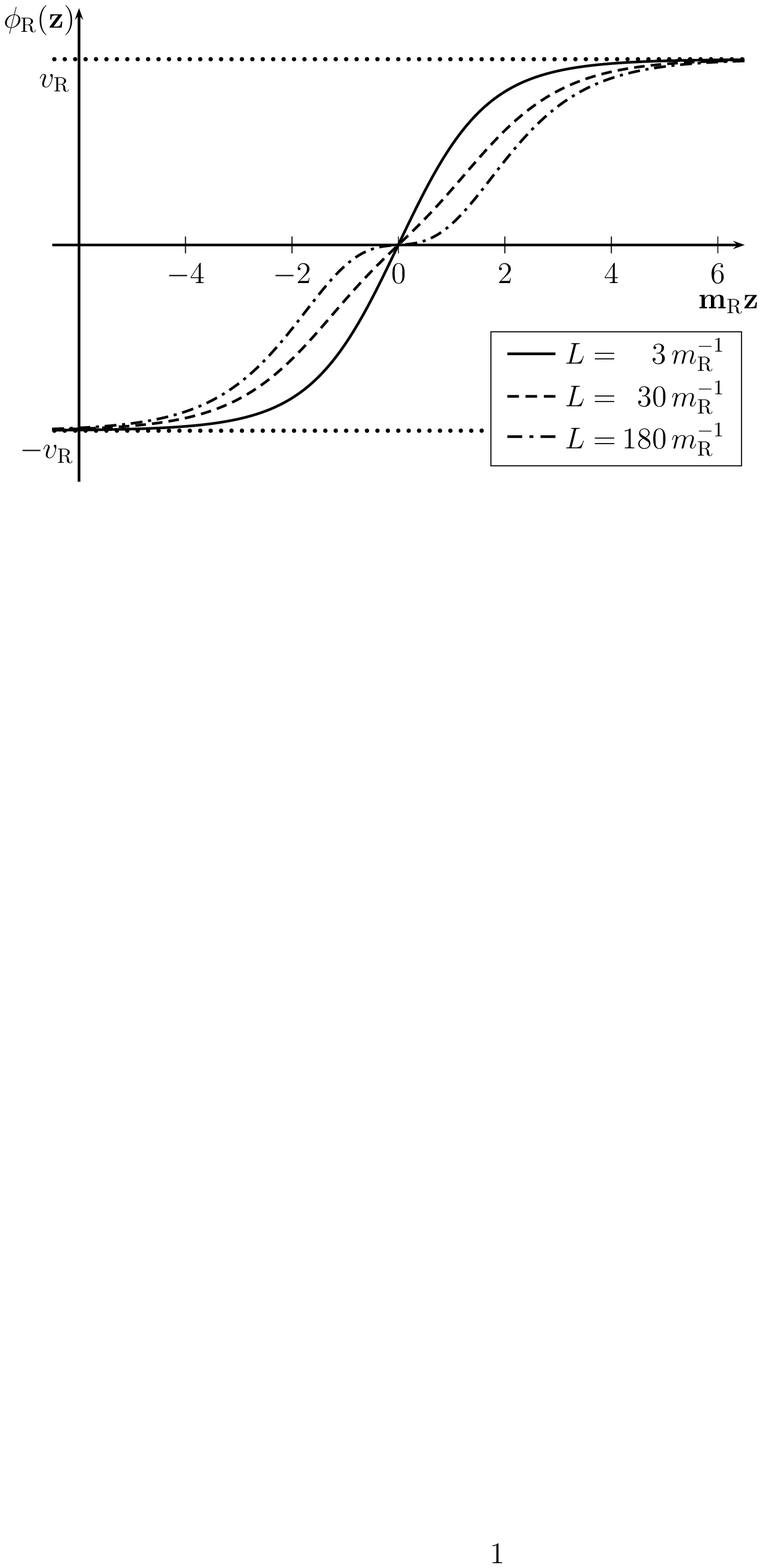}
\caption{\label{fig_profile}
\small
The renormalized interface profile for different system sizes at a coupling
of $u_R = 14.3$.}
\end{figure}

In order to illustrate the characteristics of the interface profile, we have
to specify a numerical value for the dimensionless renormalized coupling
$u_R$. In the vicinity of the critical point the coupling varies only slowly
and is close to the universal fixed point value $u_R^* = 14.3(1)$, see
\cite{CH97} for a discussion of numerical and field-theoretical estimates.
Therefore we take $u_R = 14.3$ in the plot. The interface profile according
to Eq.~(\ref{profile}) is displayed in Fig.\,\ref{fig_profile} for different
values of $m_R L$.

For $L$ larger than $200\,\xi$ the profile is no longer a monotonic function
of $z$. In this region the one-loop contribution approaches values of the
size of the leading order term and the one-loop approximation reaches its
limit of validity.

A three-dimensional calculation of the interface profile by field
theoretical methods has been performed previously by Jasnow and Rudnick
\cite{JR78} in the one-loop approximation. They do not consider systems of a
finite extent $L$, but use an external gravitational field in order to
control capillary wave fluctuations.

Their result for the interface profile has the functional form of
Eq.~(\ref{profile}), with coefficients that are given numerically. In place
of our term
\begin{equation}
- \frac{u_R}{16\pi} \ln(m_R L) \tanh(\frac{m_R}{2}z) \sech^2(\frac{m_R}{2}z)
\end{equation}
they get
\begin{equation}
c_3 \ln(h) \tanh(\frac{m_R}{2}z) \sech^2(\frac{m_R}{2}z)\,,
\end{equation}
where $h$ is the external field, and $c_3 = 0.109975$. As this term
originates from long-wavelength fluctuations, its coefficient can be
expected to correspond to ours. For a comparison one has to take into
account that $h$ corresponds to $L^{-2}$ \cite{RW82}. Also, they use an
estimate of the fixed point value of the coupling which is slightly
different from ours. In view of this, the numerical coefficient $c_3$ is in
rough agreement with our result.

The coefficients of the other terms are different from ours. We do not know,
whether there is reason to expect them to be the same.

%=======================================================================
\section{Interface width}

There are various ways to define the width $w$ of an interface, see e.g.\
refs.~\cite{Ja84}, \cite{BLS65} and \cite{MLS90}. A suitable choice is
\begin{equation}
w^2 = \langle z^2 \rangle
= \int\!\!dz\, z^2 p(z)\,,
\end{equation}
where the weight $p(z)$ is taken to be proportional to the square of the
gradient of the profile,
\begin{equation}
p(z) \propto (\partial_z \phi_R(z) )^2 ,
\end{equation}
and to be normalized:
\begin{equation}
\int\!\!dz\, p(z) = 1\,.
\end{equation}
In the evaluation of $w^2$ it should be observed that the occurring terms
have to be expanded consistently in powers of the coupling. For example, in
the one-loop approximation, the square of the gradient $\partial_z \phi =
\partial_z \phi_0 + \partial_z \phi_1$ is to be taken as
\begin{equation}
(\partial_z \phi)^2 = (\partial_z \phi_0)^2
+ 2 \partial_z \phi_0 \partial_z \phi_1
+ \mathcal{O}(u_R^2)\,.
\end{equation}
With the interface profile given above, we obtain
\begin{equation}
w^2 = \frac{b}{m_R^2}
+ \frac{3 u_R}{20 \pi m_R^2} \ln(m_R L)
\end{equation}
with
\begin{equation}
b = \frac{\pi^2 - 6}{3}
- \frac{u_R}{16\pi}\left[
\frac{12}{5} \alpha
- (\pi^2 - 6) \left( \ln 3 - \frac{13}{12} \right) \right].
\end{equation}

The interface width grows logarithmically with the system size. So the field
theoretic calculation in the one-loop approximation confirms the prediction
of capillary wave theory. Our result, however, does not rely on the
capillary wave approximation, but comes from taking into account
fluctuations of the density profile on all scales. Moreover, the numerical
coefficients are fixed unambiguously and do not depend on ad hoc cut-offs.

For a direct comparison with the convolution approximation another
definition of the interface width is more convenient, namely choosing
\begin{equation}
p(z) = (2 v_R)^{-1}\, \partial_z \phi_R(z)\,,
\end{equation}
which is meaningful as long as the profile function is monotonic. For this
choice, in the convolution approximation the squared width $\tilde{w}^2$ of
the interface equals the sum of the intrinsic and the capillary wave
contributions, see Eq.~(\ref{cawidth}).

With this definition of the interface width, our result reads
\begin{equation}
\tilde{w}^2 = \frac{a}{m_R^2} 
+ \frac{u_R}{4\pi m_R^2} \ln (m_R L)\,,
\end{equation}
with
\begin{equation}
a = \frac{\pi^2}{3} 
- \frac{u_R}{16\pi} \left\{
4 \alpha
- \pi^2 \left( \ln 3 - \frac{13}{12} \right) \right\}
= 1.249\,.
\end{equation}
By noting that for the interface tension we have \cite{Mu90}
\begin{equation}
\frac{1}{\sigma} = \frac{u_R}{2 m_R^2} + \mathcal{O}(u_R^2)\,,
\end{equation}
we see that the $L$-dependent term is in agreement with the prediction
(\ref{cwwidth}) from the capillary wave model.

The $L$-independent term $a$ contains the classical mean field value
\begin{equation}
a_0 = \frac{\pi^2}{3}\,,
\end{equation}
see \cite{MM05}, plus corrections, which are undetermined in the convolution
approximation.

For both choices of the weight function $p(z)$, the one-loop approximation
ceases to be valid, if $L$ gets so large that $p(z)$ becomes negative. This
happens for $L \approx 200\,\xi$, which coincides with the value, where the
profile begins to be non-monotonic.

Interfaces have been investigated in the three-dimensional Ising model by
means of Monte Carlo calculations in \cite{BS83,MLS90,HP92,MM05}. To observe
roughening in Monte Carlo is delicate, nevertheless one can obtain estimates
for the offset $a$ from their data, which amount to $a=2.68$ \cite{BS83},
$a=0.76$ \cite{MLS90}, $a=3.44$ \cite{HP92} and $a=0.08$ \cite{MM05}. In
view of the spread of these numbers, and in view of the fact that
higher-loop contributions will change our estimate, we can only notice that
the order of magnitude is compatible with our result.

%=======================================================================
\section{Conclusion}

Field theory, in the form of the Landau-Ginzburg model, including thermal
fluctuations on all length scales, allows to determine the interface profile
and interface width for models in the Ising universality class in the
critical region. For a system possessing a square interface of size
$L\times L$ we derived the conditional equation for the interface profile in
the one-loop approximation. We obtained its solution in analytical form.
When it is expressed in terms of physical, renormalized parameters, no
divergences occur, and there is no need to introduce ad hoc cut-offs as in
the capillary wave model.

The solution displays the characteristics of roughening by depending
logarithmically on the size $L$. The interface width grows logarithmically
with increasing system size. The coefficient of the logarithmic term is in
agreement with the universal part of the capillary wave model, and the
constant term is consistent with results from Monte Carlo simulations of the
Ising model.

%=======================================================================

\end{document}